\documentclass[11pt]{article}
\usepackage[dvips]{graphics}
\usepackage{ifthen}
\usepackage{latexsym}
\usepackage{amsmath}
\usepackage{amssymb}
\usepackage{amscd}
\usepackage{epsfig}

\def\C{{\rm\kern.24em \vrule width.02em height1.4ex depth-.05ex \kern-.26em C}}
\def\R{{\rm I\kern-.20em R}}
\def\Z{{\rm\kern.26em \vrule width.02em height0.5ex depth0ex \kern.04em
        \vrule  width.02em height1.47ex depth-1ex \kern-.34em Z}}
        \def\N{{\rm I\kern-.20em N}}
\def\Q{{\rm\kern.24em \vrule width.02em height1.4ex depth-.05ex \kern-.26em Q}}
\newcommand{\bR}{\hbox{{\rm I}\kern-.2em\hbox{{\rm R}}}}
\def\innerprod{\mathbin{\hbox to 6pt{%
                       \vrule height0.4pt width5pt depth0pt
                       \kern-.4pt
                       \vrule height6pt width0.4pt depth0pt\hss}}}

\newcommand{\Om}{{\Omega}}

\newcommand{\half}{{\mathrm{\frac{1}{2}}}}

\newcommand{\fourth}{{\mathrm{\frac{1}{4}}}}

\newcommand{\pder}[2]{{\frac{\partial #1}{\partial #2}}}
\newcommand{\vder}[2]{{\frac{\delta #1}{\delta #2}}}

\newcommand{\inner}[2] {{i(#1)\,#2}}
\newcommand{\Lie}[2] {{{\mathcal L}_{#1} #2}}
\newcommand{\W}{{\,\wedge\,}} 

\newcommand{\dsp}{\displaystyle}

\newcommand{\bc}{\begin{center}}
\newcommand{\ec}{\end{center}}

\newcommand{\Ham}{{\mathcal H}}
\newcommand{\bHam}{{\mathbf \Ham}}
\newcommand{\Lam}{{\Lambda}}
\newcommand{\bLam}{{\mathbf \Lam}}

\newcommand{\bH}{{\mathbf H}}

\newcommand{\bS}{{\mathbf S}}
\newcommand{\bW}{{\mathbf W}}

\newcommand{\bK}{{\mathbf K}}
\newcommand{\bM}{{\mathbf M}}
\newcommand{\bz}{{\mathbf z}}

\newcommand{\Svec} {{\vec{S}}}

\newcommand{\Sb}{{\overline S}}
\newcommand{\wb}{{\overline w}}
\newcommand{\zb}{{\overline z}}

\newcommand{\wsq}{{|w|^2}}

\newcommand{\p}{{p}}
\newcommand{\pb}{{\overline p}}
\newcommand{\psq}{{|p|^2}}

\newcommand{\w}{{w}}
\newcommand{\wt}{{w_t}}
\newcommand{\wx}{{w_x}}
\newcommand{\wxx}{{w_{xx}}}
\newcommand{\wbt}{{\wb_t}}
\newcommand{\wbx}{{\wb_x}}
\newcommand{\wbxx}{{\wb_{xx}}}
\newcommand{\px}{{\p_x}}
\newcommand{\delp}{{\delta \p}}
\newcommand{\pbx}{{\pb_x}}
\newcommand{\delpb}{{\delta \pb}}

\newcommand{\dw}{{\delta \w}}
\newcommand{\dwb}{{\delta \wb}}
\newcommand{\dwx}{{\delta \wx}}
\newcommand{\dwbx}{{\delta \wbx}}
\newcommand{\dz}{{\delta z}}
\newcommand{\dzb}{{\delta \zb}}
\newcommand{\Sone}{{S_1}}
\newcommand{\Stwo}{{S_2}}
\newcommand{\Sthree}{{S_3}}

\newcommand{\Sonex}{{S_{1,x}}}
\newcommand{\Stwox}{{S_{2,x}}}
\newcommand{\Sthreex}{{S_{3,x}}}

\newcommand{\wone}{{w^1}}
\newcommand{\wtwo}{{w^2}}
\newcommand{\wthree}{{w^3}}
\newcommand{\wfour}{{w^4}}
\newcommand{\woneb}{{\wb^1}}
\newcommand{\wtwob}{{\wb^2}}
\newcommand{\wthreeb}{{\wb^3}}
\newcommand{\wfourb}{{\wb^4}}

\newcommand{\wbtil}{{\tilde \wb}}
\newcommand{\ab}{{\overline a}}
\newcommand{\bb}{{\overline b}}
\newcommand{\vb}{{\overline v}}
\newcommand{\ub}{{\overline u}}
\newcommand{\Rt}{{\tilde R}}
\newcommand{\Lh}{{\hat L}}
\newcommand{\Kh}{{\hat K}}
\newcommand{\Jh}{{\hat J}}

\newcommand{\cref}[2] {{(\ref{#2})}}

\newcommand{\Sec} {{\mathcal S}}
\newcommand{\dLam} {{\tilde \Lambda}}
\newcommand{\tbLam} {{\tilde \bLam}}
\newcommand{\Nr} {{N^{(r)}}}
\newcommand{\ua} {{u^{(\alpha)}}}
\newcommand{\uaj} {{u^{(\alpha)}_j}}
\newcommand{\us} {{u_\Sigma}}

\newcommand{\dxn} {{dx_1 \wedge \dots \wedge dx_n}}
\newcommand{\dxni} {{(-1)^{i-1} dx_1 \wedge \dots \wedge \widehat dx_i \wedge \dots \wedge dx_n}}

\newcommand{\Sigmar} {{\Sigma^{(r)}}}
\newcommand{\sigmar} {{\sigma^{(r)}}}

\newcommand{\pdx} {{\frac{\partial}{\partial x}}}
\newcommand{\pdxi} {{\frac{\partial}{\partial x_i}}}
\newcommand{\pdi}  {{\partial_i}}
\newcommand{\pduj} {{\frac{\partial}{\partial u_j}}}     
\newcommand{\pduaj}{{\frac{\partial}{\partial u^{(\alpha)}_j}}} 
\newcommand{\ddt} {{\frac{d}{dt}}}
\newcommand{\Ninf} {{\overline N}}
\newcommand{\piu}[1]{{\pi^{(#1)}}}

\newcommand{\Ainf}{{\underline A}}
\newcommand{\dDR} {{\underline d}}
\newcommand{\Hor}{{\text{Hor}}}
\newcommand{\Ver}{{\text{Ver}}}
\newcommand{\Zbar} {{\overline Z}}

\newcommand{\Wbar} {{\overline W}}
\newcommand{\cH} {{\mathcal H}}
\newcommand{\Omegai} {\Omega^{(i)}}

\newcommand{\Thetai} {\Theta^{(i)}}
\newcommand{\Thetaj} {\Theta^{(j)}}
\newcommand{\Thetak} {\Theta^{(k)}}

\title{Local Lagrangian Formalism and Discretization of the Heisenberg Magnet Model}

\author{D. Karpeev\\
Mathematics and Computer Science Division\\
Argonne National Labs\\
karpeev@mcs.anl.gov\\
 and\\
  C.M. Schober\thanks{This work was partially supported by the NSF, grant number DMS-0204714.}\\
 Department of Mathematics\\
University of Central Florida\\ cschober@mail.ucf.edu 
}
\date{}
\begin{document}
\maketitle
\abstract
In this paper we develop the Lagrangian and
multisymplectic structures of the Heisenberg magnet (HM) model 
which are then used as the basis for geometric discretizations of HM.
Despite a topological obstruction to the existence of a global 
Lagrangian density, a local variational 
formulation allows one to derive local conservation laws using a 
version of N\"other's theorem from the formal variational 
calculus of Gelfand-Dikii.
Using the local Lagrangian form we extend the method of Marsden, Patrick and Schkoller to derive local multisymplectic 
discretizations directly from the variational principle.  We employ a version of the finite element method to discretize
the space of sections of the trivial magnetic spin bundle $N = M\times S^2$ over an appropriate space-time $M$. 
Since sections do not form a vector space, the usual FEM bases can be used only locally with coordinate transformations 
intervening on element boundaries, and conservation properties 
are guaranteed only within an element.  We discuss
possible ways of circumventing this problem,  including the 
use of a local version of the method of characteristics, 
non-polynomial FEM bases and Lie-group discretization methods.

\section{Introduction}
The treatment of PDEs in invariant form finds its most natural language in the setting of jet bundles over an appropriate
space-time.  In this setting, governing equations of ``motion'' are 
prescribed by invariantly-defined differential operators, and
although calculations are ultimately done in terms of partial derivatives, 
the invariant meaning of all the operations is
clear. Moreover, the invariant formulation and structure of the underlying fiber bundle frequently translate to
a nontrivial geometry of the space of solutions, conservation properties and other ``geometric'' features.  Although
these geometric properties can be discovered in the coordinate formulation, conceptual clarity of 
invariant geometric derivations is frequently a great asset in itself.  

All of this is particularly so for Langrangian PDEs whose equations of motion are derived as extermality conditions
on the action functional -- an apparently coordinate-free form.  Symplectic and multisymplectic structures immediately 
follow by well-known procedures, as do conserved quantities corresponding to continuous symmetries (N\"other's theorem).
The situation is less fortunate for Hamiltonian systems as here a number of ``symmetries'' have been broken relative to
the Lagrangian setting. A splitting of space-time relies upon the {\em choosing}  a preferred time direction and 
a nonunique 
complementary space -- a literal symmetry breaking as the full group of space-time transformations no longer preserves
the form of the equations.  Furthermore, the system is no longer specifed by 
prescribing a single quantity -- apart from the
energy functional a Poisson structure has to be specified independently\footnote{
In fact, there is a certain nonuniqueness associated with the choice of the {\em canonical} 1-form in Lagrangian theories
living on jet bundles $\Nr$ with $r > 1$ \cite{deligne+.I, manin}, although this is not as bad as the number of choices that need to be made
in the Hamiltonian setting.}.  

Although it may appear from familiar examples of classical mechanics that the 
Lagrangian and Hamiltonian pictures are equivalent, this is not always so.  In fact there is a local nondegeneracy
condition on the Poisson bracket that must be satisfied before a {\em local} Lagrangian density 
can be reconstructed. In
general, an additional vanishing condition on a topological obstacle 
(i.e. exactness of the symplectic form) has to be satisfied 
before the variational picture is restored globally.  Even so, a local Lagrangian formulation carries enough
information to reconstruct the local multisymplectic geometry and N\"other-type local conservation laws.  
A central result is the derivation of the local multisymplectic form 
for the Heisenberg Magnet model from its local Lagrangian formulation.

Even when Lagrangian and Hamiltonian theories are equivalent, their discretizations may not be so.  
While the discrete
variational principle yields analogs of symplectic and multisymplectic geometries under minimal conditions on the
discretization, the Hamiltonian case calls for a sophisticated theory of symplectic or Poisson integrators to preserve
the corresponding geometries.  Therefore, the search for an equivalent Lagrangian formulation naturally arises
when considering conservative discretizations.  In this paper we discuss potential ways of constructing geometric
discretizations from the {\em local} variational picture.  

In Section~\ref{Section.Setting} we briefly recall the picture of
geometric PDEs as conditions on sections of an appropriate fiber bundle and their jets.  The formal calculus of
variations of Gelfand-Dikii is a very convenient computational tool for Lagrangian systems in this setting,  
and  a source of many explicit formulas, including local conserved densities and a formal analog of N\"other's theorem.  
Section~\ref{Section.HM-Lagrangian} contains the derivation of the local Lagrangian form of HM, its conservation laws, and
topological obstructions to the existence of a global description. 
The relation to Novikov's multivalued version of Morse
theory as well as  Lyusternik-Schnirelmann theory are discussed in this section.  In Section~\ref{Section.HM-Discr} we discuss
variational discretizations of HM using a version of the finite-element method.  
An essential difficulty here is that the sections 
are constrained to lie on the unit 2-sphere, which is not satisfied by the standard FEM bases, and the
resulting appearence of rational terms in the discrete Lagrangian.  
We discuss ways of avoiding this problem, including the use of discretizations based on Lie group methods to
satisfy the sphere constraint in the most natural
manner.  


\section{Setting}
\label{Section.Setting}

The geometric theory of differential operators and variational calculus is
most naturally set within the framework of smooth fiber bundles $N \xrightarrow{\pi} M$  over some
$n-$dimensional base space-time $M$ and the corresponding jet bundles $\Nr \xrightarrow{\piu{r}} M$
\cite{steenrod}.
Differential operators and action functionals of 
the Lagrangian formalism act on the space of $\Sec(N)$ sections of the bundle,
each section  $\sigma: M \rightarrow N$, $\pi \circ \sigma = \text{id}_M$ identifiable with its  image $\Sigma =
\sigma(M)$, a copy of $M$ horizontally embedded in $N$.
Fixing a local trivialization $N {\cong}_{\text{loc}} M \times F$
expresses $N$ as a product of $M$ with the $k$-dimensional typical fiber $F$.
A choice of local coordinates, $x = (x_i), i = 1, \dots, n$ on the base and $u = (u_j), j = 1,\dots, r$ 
on the the fiber, defines a local coordinate system 
$x\times u: N \rightarrow \R^{n+k}$, making $N$
locally diffeomorphic to the canonical projection $\pi \cong \R^{n+k} \rightarrow \R^n$, while 
inducing a similar diffeomorphism $\piu{r} \cong \R^{n+k} \times \R^{n_r} \rightarrow \R^n$ for the $\Nr$ with 
the {\em adapted} coordinates $\left((x_i),(\uaj)\right), |\alpha| \leq r$\footnote{
        Definitions for multi-indices are as usual: $\alpha = (\alpha_1, \dots, \alpha_n),\ |\alpha| = \sum_i \alpha_i$,
        $\left(\pdx\right)^{\alpha} \equiv \frac{\partial}{\partial x_1}^{\alpha_1} \cdots \frac{\partial}{\partial x_n}^{\alpha_n}$,
        $\partial^\alpha = \partial_1^{\alpha_1} \cdots \partial_n^{\alpha_n}$, $\epsilon_i$ has $1$ in
        the i-th place and zeros everywhere else.
}
on $\Nr$.  The usual bundles $\wedge^k TN$  and $\Lambda^k N$ have as their spaces of sections 
the $k$-vector fields $V^k(N) = \Sec(TN)$ and the differential $k$-forms $\Omega^k N = \Sec(\Lambda^k
N)$.  In addition, there is a distinguished subspace $V(N,M) \subset V(N) \equiv V^1(N)$ of fiber-preserving fields,
which generate local fiber bundle automosphisms,
and the subspace $V(N/M) \equiv \Sec(T(N/M)) \subset V(N,M)$ of sections of the vertical subbundle, which fix fibers.
When restricted to $\Sigma$  the local coordinates $u$ define functions of $x$ by 
$$ 
        \us: \R^{n} \xrightarrow{x^{-1}} M \xrightarrow{\sigma} \Sigma \xrightarrow{u}\R^k.
$$
Similarly, restriction of the fiber coordinates $\ua$ to $\Sigmar$ defines functions $u_\Sigmar^{(\alpha)}$
that encode the infinitesimal behavior of the section: $\left(\pdx\right)^\alpha \us(x) = u_\Sigmar^{(\alpha)}(x), \quad |\alpha| \leq r$.

To the variational principle we admit {\em local} functionals $\tbLam$ that to each $\sigma \in \Sec(N)$ assign the integral
$\int \dLam(\sigma)$ of an $n$-form $\dLam(\sigma) \in \Omega^n M$,
such that the integrand $\dLam_x(\sigma) \in \Lambda_x^n M$ at any $x \in M$ is determined by the $r$-jet
$\sigma^{(r)}_x$ of the section at that point\footnote{
        For a globally determined $r$.
}.  
In other words $\tbLam$ is a differential operator $\Sec(N) \rightarrow \Omega^n M $ of order $\leq r$ and by definition
can be factored through a map $\bLam_*$ induced on sections by a smooth bundle map $\bLam: \Nr \rightarrow \Lambda^n M$ over $M$. 
$$
        \begin{CD}
               \Sec(\Nr) @>{\bLam_*}>>          \Omega^n M\\
                @A{J^r_*}AA                       @VV{\int}V\\
                \Sec(N)    @>{\tbLam}>>          \R
        \end{CD}.
$$
By a certain abuse of notation defined in \cite{palais} we shall call $\bLam$ the {\em symbol} of the operator $\tbLam$
and write 
$$
        \tbLam(\sigma) = \int \bLam_*(\sigmar),
$$
where in order to make sense of the integral we restrict integration to be over {\em nice} (see
\cite{marsden+patrick+shkoller}) compact domains $\Sigma_M
\subset M$.

The symbol $\bLam$ is identified with a section of the {\em vector} subbundle 
$\Lambda^{0,n} \Nr \equiv \piu{r}^* \Lambda^n M \subset \Lambda^n \Nr$ of horizontal forms vanishing on the fibers of the vertical subbundle $T(\Nr/M)$.  
Then the action of $\tbLam$ can be defined simply by integration of its symbol over holonomic\footnote{
        In contrast to $N^{0} \cong N$, for $r > 0$ not every horizontal submanifold $\Sigma_r \subset \Nr$ is a lift of some
        $\Sigma$, since $\Sigmar$ must satisfy the holonomy
        constraint $\left(\pdx\right)^\alpha u_\Sigmar^{(\beta)}(x) = u_\Sigmar^{(\alpha + \beta)}(x)$. 
}  horizontal
submanifolds $\Sigmar = \sigmar(\Sigma_M) \subset \Nr$:
\begin{equation}
        \tbLam_\Sigma \equiv \tbLam(\sigma) = \int_{\Sigmar} \bLam.
\label{caf}
\end{equation}
In terms of this geometric description, variational calculus investigates the response of $\tbLam$ to
infinitesimal variations of the holonomic submanifold $\Sigmar$.

A generic variation $Z_\Sigma$ of $\Sigma$ is represented by a (local) $1$-parameter bundle automorphism group $\phi^t_N$, with the induced (local) base automorphism
group $\phi^t_M = \pi \circ \phi^t_N$,  by restriction $Z_\Sigma = Z|_\Sigma$ of the infinitesimal generator $Z \in V(N,M)$.
In coordinates we have
\begin{equation}
        \label{setting.Z}
               Z = \left.\frac{d}{dt}\right|_{t=0} {\phi^t_N}^* = X(x) + V(x,u) = \sum_i X_i(x) \pdxi + \sum_j V_j(x,u) \pduj.
\end{equation}
The action of $Z$ on $\Omega N$ naturally splits into the horizontal $X$ and vertical $Y$ parts.
The horizontal part acts on functions $A_0 = C^{\infty} (N)$ by differentiation along $\Sigma$, i.e. via pushing
$\pdxi$ forward to $\Sigma$, while the
vertical part acts by differentiating along the vertical curves $\sigma^t(x)$ via the natural action of $\phi^t_N$ sections
$\sigma^t = \phi^t_N \circ \sigma \circ \phi^{-t}_M$:
$\dsp{X(x,\us(x)) = \sigma_* \pdxi = \sum_i X_i \left(\pdxi + \sum_j\us_{j,i} \pduj \right)}$,\\
$\dsp{Y(x,\us(x)) = \sum_j \ddt \us^t_j(x) \pduj = \sum_j \left( V_j(x,\us(x)) - \sum_i \us_{j,i}(x) X_i(x)\right) \pduj}$.\\
Geometrically, $Y$ is the projection of $Z$ onto the vertical subspace at $T_{\sigma(x)} (N/M)$ {\em parallel} to the horizontal 
subspace $T_x \Sigma = T_x\sigma \cdot T_x M$, and $X$ is the projection {\em onto} $T_x \Sigma$.
 
The action of $Z$ on the higher jets $\sigmar(x)$ is easily computed by interchanging the order of differentiation, for instance,
\begin{multline*}
        Y^{(1)} = Y  + \sum_{j,i} \ddt \pdxi \us^t_j(x) \frac{\partial}{\partial u_{j,i}} 
                 = Y  + \sum_{j,i} \pdxi \ddt \us^t_j(x) \frac{\partial}{\partial u_{j,i}} 
                 = Y  + \sum_{j,i} \pdxi Y_j(x,\us(x))  \frac{\partial}{\partial u_{j,i}}.
\end{multline*}
Thus, the first jet components of $Y^{(1)}$ are obtained from components of $Y$ by horizontal differentiation along $T_x
\Sigma$, using the second jet information
$\us_{j,i^\prime i}(x)$ to determine the horizontal subspace $T \Sigma^{(1)}$.  The general rule follows recursively.

Jet bundles allow one to abstract from particular sections by encoding their infinitesimal information pointwise, 
similar to  the way tangent bundles allow one to manipulate vectors without referring to local integral curves. 
In general,
$\pdxi$ will act on algebras of function $f \in A_r = C^\infty(\Nr)$ by the {\em horizontal holonomic lift}
$\pdi = \Hor(\pdxi)$ and  has the property that it  reduces to the total derivative $\frac{d}{d x_i}$ 
upon restriction to a holonomic section: $\pdi = \sigmar_* \pdxi$.
Since, as seen in the example above, $\partial_i$ uses the higher jet data, it acts on $A_r$ as a map to $A_{r+1}$: on coordinates we have
$\partial_i \uaj = u_j^{(\alpha + \epsilon_i)}$.  
Therefore the operators $\partial_i$ must be regarded as derivations of
the algebra $\Ainf = \injlim A_r \cong \bigcup  A_r$ of functions on the infinite jet space $\Ninf = \projlim \Nr$.
All derivatives $\text{Der}(\Ainf)$, formally viewed as vector fields $V(\Ninf)$ on $\Ninf$ are given by the infinite series 
$\overline W = \sum_i W_i \pdxi + \sum_{j,\alpha} W_{j,\alpha} \pduaj$, and 
derivations $\partial_i$ lie within the subspace $V(\Ninf, M)$ of formally fiber preserving fields, 
thus generating the
submodule $\Hor(\Ninf)$ over $K$.  Likewise the {\em vertical holonomic lifts} $\Ver(Y)$ of fields on $Y \in V(N/M)$
generate $\Ver(\Ninf) \in V(\Ninf/M) \subset V(\Ninf,M)$ as a module over $\Ainf$:
$$
        \partial_i = \pdxi + \sum_{j,\alpha} u_j^{(\alpha + \epsilon_i)} \pduaj,\quad 
        \Ver(Y) = \sum_{j,\alpha} Y_j^{(\alpha)} \pduaj \in V(\Ninf, M), \quad Y_j \in A_0.
$$
It is easily checked that $\Ver(\Ninf)$ commutes with all $\pdi$.  

Geometrically, fields in $\Hor(\Ninf)$ reduce to
$V(M)$ and fields in $\Ver(\Ninf)$ reduce to $V(N/M)$ upon restriction to
 holonomic submanifolds $\Sigmar \subset \Nr \subset \Ninf$.
In particular, the above $Z$, when regarded in $V(\Ninf, M)$ uniquely splits into 
a ``horizontal'' and ``vertical''
part,  
\[\dsp{X = \sum_i X_i \left(\pdxi + u_{j,i} \pduj\right)\dsp},\qquad
\dsp{Y = \sum_j Y_j \pduj = \sum_j\left(V_j - \sum_i u_{j,i}\right) \pduj}.\]
Each part uniquely
lifts to $\overline X = \Hor(\overline Z) \in \Hor(\Ninf)$ and $\overline Y = \Ver(\overline Z) \in \Ver(\Ninf)$,
and together define 
a unique lift $\overline Z = \Hor(\overline Z) + \Ver(\overline Z) \in \Hor(\Ninf) \oplus \Ver(\Ninf)$.

The direct sum $\Hor(\Ninf) \oplus \Ver(\Ninf) \subset V(\Ninf, M)$, 
generates a split of the space of $1$-forms 
$\Omega^1 \Ninf = \injlim \Omega^1 \Nr$ and of the usual ``de Rham'' differential
$\dDR$ into the vertical and horizontal parts: $\Omega^1 \Ninf = \Omega^{1,0} \Ninf \oplus \Omega^{0,1} \Ninf$ and $\dDR
= \delta + d$ respectively, where $\Omega^{1,0} \Ninf = \delta \Ainf$ consists of
forms vanishing upon contraction with horizontal fields, and $\Omega^{0,1} \Ninf = d \Ainf$ vanishing upon contraction
with vertical vector fields.  The geometric meaning of the horizontal and vertical differentials is that $d$ represents
differentiations along the horizontal subspace at each level $\Sigmar$and  reduces to the differential in
$\Omega^l M$ upon restriction to a holonomic section, while $\delta$ vanishes, measuring the ``virtual
variations'' transversal to holonomic $\Sigmar$.

Operating formally we can deduce the properties of the calculus defined on $\Omega \Ninf$ by the pair of differentials \cite{dikii}.
In particular, $\delta$ and $d$ anti-commute with one another, and on the generators we have
\begin{gather*}
                dx_i = \dDR x_i,\  d\uaj = \sum_i u_j^{(\alpha+\epsilon_i)} dx_i, \  df = \sum_i \pdi f dx_i,\\
                \delta \uaj = \dDR \uaj - \sum_i u_j^{(\alpha + \epsilon_i)} dx_i,\ \delta x_i = 0,\ 
                \delta f = \sum_{j,\alpha} \pder{f}{\uaj} \delta \uaj,
\end{gather*}
and the module $\Omega \Ninf$ decomposes with respect to $\delta$ and $d$
\begin{gather*}
        \Omega^m \Ninf = \sum_{k+l=m} \Omega^{k,l} \Ninf = \sum_{k+l=m} \Omega^{k,0} \Ninf \wedge \Omega^{0,l} \Ninf,\quad
        \delta: \Omega^{k,l} \Ninf \rightarrow \Omega^{k+1,l},\quad 
        d:\Omega^{k,l} \rightarrow \Omega^{k,l+1}
\end{gather*}
into exterior subalgebras of vertical $k$-forms $\Omega^{k,0} \Ninf = \Ainf[\delta u_{j_1}^{(\alpha_1)}, \dots, \delta u_{j_k}^{(\alpha_k)}]$,
and horizontal $l$-forms \\$\Omega^{0,l} \Ninf \Ainf[dx_{i_1}, \dots, dx_{i_l}]$ (in particular, $\Omega^l \Ninf = 0$ for $l >
n$), with each subalgebra $\Omega^{k,l} \Ninf$ filtered $\Omega^{k,l}_0 \Ninf \subset \cdots \subset
\Omega^{k,l}_r \Ninf \subset \cdots $ by the subalgebras of $k,l$-forms  
$\Omega_r^{k,l} \Ninf = \Ainf[\delta u_{j_1}^{(\alpha_1)}, \dots, \delta u_{j_k}^{(\alpha_k)},dx_{i_1}, \dots,
dx_{i_l}]$  of 
{\em differential degree} $\text{deg} = \sum_s |\alpha_s| \leq r$.

The fields $V(\Ninf)$ act on $\Omega \Ninf$ by Lie derivatives 
$\Lie{\Wbar}{\Omega} \equiv \Wbar (\Omega) = (\dDR \inner{\Wbar} + \inner{\Wbar}\dDR) \Omega$, 
with the only nonvanishing contractions being $\inner{\pdi} dx_i = 1$ and $\inner{\pduaj} \delta \uaj = 1$.
The easily observed commutation relations 
simplify calculations of $\Lie{\Ver(\Ninf)}{}$ and $\Lie{\Hor(\Ninf)}{}$:
\begin{gather*}
        \delta \inner{\Hor(\Ninf)} + \inner{\Hor(\Ninf)} \delta = d \inner{\Ver(\Ninf)} + \inner{\Ver(\Ninf)} d = 0,\\
        \Zbar(\Omega) = 
        \left(\delta \inner{\Ver(\Zbar)} + \inner{\Ver(\Zbar)} \delta\right)\Omega + 
        \left(d \inner{\Hor(\Zbar)}{} + \inner{\Hor(\Zbar)}{} d \right)\Omega.
\end{gather*}
This forms the basis of formal variational calculus, developed in detail in \cite{dikii, manin},
here we sketch the necessary results.

The main result is the formula for the variational derivative (\ref{setting.var}), 
which computes the action of $Z$ in (\ref{setting.Z}) on the
symbol of the action functional 
$\bLam \in \Omega^{0,n} \Nr$ via the holonomic lift:
$$
        Z(\bLam) \equiv \Zbar(\bLam) = \inner{\Ver(\Zbar)} \delta \bLam + d \inner{\Hor(\Zbar)} \bLam,
$$
since $\inner{\Ver(\Ninf)} \Omega^{0,l} = d \Omega^{k,n} = 0$ for any $k,l$.  This shows in what sense the vertical part
of $\overline Y$ acts on $\bLam$ as on a function, and the horizontal part $\overline X$ acts on $\bLam$ as on a horizontal $n$-form (see
\cite{marsden+patrick+shkoller}, eq. (4.37)).  In coordinates we have 
\begin{multline*}
        \bLam = \Lam \dxn,\ \Lam \in \Ainf, \quad
         \delta \bLam = \sum_{j,\alpha} \pder{\Lam}{\uaj} \delta \uaj \wedge \dxn = \\
        \sum_{j,\alpha} \pder{\Lam}{\uaj} \,\delta(\partial^\alpha u_j) \wedge \dxn.
\end{multline*}
By a formal differential analog of integration by parts we can split $\delta \bLam$ as follows:
\begin{multline}
        \label{setting.var}
        \delta \bLam = 
                        \delta u_j \underbrace{\vder{\Lam}{\uaj} \wedge \dxn}_{\vder{\bLam}{u_j} \in \Omega_0^{0,n}} - \\
                        d\left(\sum_i \Theta_i \dxni\right)  \in \Omega_0^{1,n} \Ninf\, \oplus\, d\Omega^{(1,n-1)} \Ninf.
\end{multline}
The geometric meaning of the variational derivative
operator is that it provides a  formal adjoint to the operator of vertical lift on
vectors $Y \in V(N/M)$, modulo the horizontal differential $d$:
\begin{equation}
        \inner{\Ver(Y)} \delta \bLam = \inner{Y} \delta \bLam_0 - \inner{Y} d \Theta = \inner{Y} \delta \bLam_0 + d \left( \inner{Y} \Omega^{(1)}\right),
\label{formadj}
\end{equation}
where $\delta \bLam_0 \in \Omega^{1,n}_0 \Ninf$ is the variational derivative.
Integrating (\ref{formadj})
over some $\Sigmar$, since  $Y$ vanishes on the boundary of $\partial \Sigma$, the adjoint meaning becomes
exact and the  {\em equations of motion} defining the extrema of $\tbLam$ are given locally by
\begin{equation}
        \label{setting.EL}
        \vder{\bLam}{u_j} = 0.
\end{equation}

Define the {\em multisymplectic form} $\Omega$ and the vertical part $X_i$ of $\partial_i$ by
\begin{equation}
        \label{setting.Omega}
        \Omega = \delta \Theta \in \Omega^{2,n-1},\ \Omegai = \delta \Thetai \in \Omega^{2,0},\quad
        X_i = \partial_i - \pdxi = \sum_{j,\alpha} u_j^{(\alpha+\epsilon_i)} \in V(\Ninf/M).
\end{equation}
On any holonomic $\Sigmar$ satisfying (\ref{setting.EL})
we have the equivalent {\em covariant Hamiltonian} system, which, when $\bLam$ does not depend on the base
coordinates $x$ explicitly, has the form\footnote{
        Which depends on the choice of trivialization.
}
\begin{equation}
        \label{setting.cov}
        \sum_i \inner{X_i} \Omegai = -\delta \cH, \quad \cH = \sum_j \inner{X_i} \Thetaj - \Lam \in \Ainf,
\end{equation}
obtained by a version of the Legendre transform.
Among other things, it implies the {\em preservation of the multisymplectic form}
\begin{equation}
        \label{setting.ms}
        \sum_i X_i(\Omegai) = -\delta^2 \cH = 0,
\end{equation}
as well as the local conservation laws (formal N\"other's theorem):
\begin{gather}
        \label{setting.conserv}
        \sum_k \partial_k T_{jk} = 0,\quad
        T_{jj} = \cH - \sum_{k \neq j} \inner{X_k}{\Thetak},\ T_{jk} = \inner{X_j}{\Thetak}, j \neq k. 
\end{gather}
Later we will have a chance to use the {\em momenta} defined as \cite{dikii}
\begin{equation}
        \label{setting.momenta}
        P_{j,(\alpha)} = \vder{\bLam}{\uaj} = 
        \sum_{\beta} (-1)^{|\beta|} \frac{|\beta|}{|\alpha|} 
        \frac{(\alpha_1 + \beta_1) \cdots (\alpha_n + \beta_n)}{\beta_1 \cdots \beta_n} 
        \frac{\partial^{\beta} \bLam}{u_j^{(\alpha + \beta)}},
\end{equation}
in particular, $P_{j,(0)} = \vder{\bLam}{u_j}$.

\newcommand{\bomega}{{\mathbf \omega}}
\newcommand{\btheta}{{\mathbf \theta}}
\newcommand{\tbomega}{{\tilde \bomega}}
\newcommand{\tbtheta}{{\tilde \btheta}}
\newcommand{\tbH}{{\tilde \bH}}

The connection with the usual time formalism is done as follows \cite{dikii}.  Fix a space-time split of $M = M^{n-1}
\times M^1$ with a compatible coordinate system $(x_1, \dots, x_{n-1}, t = x_n)$.  For any section $\Sigma$
fix the {\em time slice} $\Sigma(t^\prime)$ over the hyperplane $M^{n-1}\times {t}$ defined by $x_n = t$.  
For the holonomic lift $\Sigmar(t)$ of any such slice, we define the {\em symplectic form} $\tbomega$, 
 the corresponding primitive
$1$-form $\tbtheta$ and the Hamiltonian functional $\tbH$ with the corresponding densities
\begin{equation}
        \tbomega(\sigma) = \int_{\Sigmar(t)} \bomega,\quad
        \tbtheta(\sigma) = \int_{\Sigmar(t)} \btheta,\quad
        \tbH(\sigma) = \int_{\Sigmar(t)} \underbrace{T_{nn} dx_1 \wedge \dots \wedge dx_{n-1}}_{\bH},
\end{equation}
provided the Lagrangian $\bLam$ is {\em regular} \cite{dikii, marsden+patrick+shkoller}.
Then the Hamiltonian evolutionary equations are given as usual by 
$$
        \delta \tbH = \inner{X_t} \tbomega,
$$
and the Lagrangian density is recovered by a version of the inverse Legendre transform:
\begin{equation}
        \label{setting.lagr}
        \bLam = (-\bH + \inner{X_t} \btheta)\wedge dx_n.
\end{equation}

\section{Local Lagrangian of HM System}
\label{Section.HM-Lagrangian}

\newcommand{\Sigmat}{{\Sigma(t)}}
\newcommand{\sigmat}{{\sigma(t)}}
\newcommand{\Sigmatk}[1]{{\Sigma^{(#1)}(t)}}
\newcommand{\sigmatk}[1]{{\sigma^{(#1)}(t)}}
\newcommand{\Nt} {{N(t)}}
\newcommand{\Ntk}[1] {{N^{(#1)}(t)}}
\newcommand{\bF}{{\mathbf F}}
\newcommand{\bG}{{\mathbf G}}
\newcommand{\tbF}{{\tilde \bF}}
\newcommand{\tbG}{{\tilde \bG}}
\newcommand{\Func}{{\mathcal F}}
\newcommand{\uvec}{{\vec{u}}}
\newcommand{\Svect}{{\Svec(t)}}
\newcommand{\sothree}{{\text{so}(3)}}
\newcommand{\SOthree}{{\text{SO}(3)}}

In this section 
we derive a local Lagrangian form of the action functional  of the Heisenberg magnet (HM) model
since the symplectic form corresponding to the Poisson 
bracket of HM is not exact and  a global Lagrangian does not exist.  
Throughout this section we explicitly work in coordinates.

The HM model is a Hamiltonian system
defined on the product fiber bundle $N = M \times \R^3$, where the space-time base is
simply $M = \R^2$ with coordinates $(x_1, x_2)$ on the base and $\Svec = (S_1, S_2, S_3)$ on the fiber
(e.g. \cite{faddeev+takhtajan}).  A solution of the
system is a section $\Sigma \in N$ represented by the restriction of the fiber coordinates 
$\Svec : M \rightarrow \R^3$ and satisfying the following equation 
$$
        \partial_{2} S_a - \sum_{b,c} \epsilon_{a b c} S_b \partial_{1}^2 S_c = 0.
$$
We now reconstruct the Hamiltonian structure of HM in two different coordinate systems on the fiber.

\subsection{Hamiltonian form of HM}
\label{SubSection.HM-H}
After the space-time split $M \cong M^{n-1} \times M^1 = \R \times \R$ with coordinates $x = x_1$ and $t = x_2$,
we have the instantaneous state of the system $\Sigma(t) = \sigma(t)(M^{n-1})$ 
at time $t \in M^1$ given by a section $\sigmat$ of the instantaneous bundle $N(t) = M^{n-1} \times \R^3 \rightarrow M^{n-1}$ and 
represented by a function $\Svect: M^{n-1} \rightarrow \R^3$. THe notation from Section~\ref{Section.Setting} is modified
accordingly: $A_r(t), \Ainf(t), \Omega \Nr(t)$ etc.
The Hamiltonian functional of the system is
$$
  \label{hml.Ham}
   \tbH(\sigmat) = \int_{\Sigmatk{1}} \half \left(\Sonex^2 + \Stwox^2 + \Sthreex^2\right) dx,
$$
prescribed by the horizontal $1$-form  $\bH =  H dx = \half \left(\Sonex^2 + \Stwox^2 + \Sthreex^2\right) dx \in
\Omega^{0,1} \Ntk{1} \in \Omega^{0,1} \Ninf(t)$ with the
density $H \in A_1(t)$.  There is a local Poisson bracket on each fiber defined in coordinates by
\begin{equation}
  \label{hml.bracket}
        \{S_a, S_b\} = \Psi_{a,b}(\Svec) = -\sum_c \epsilon_{a b c} S_c,
\end{equation}
which can be viewed as a section over $\Nt$ of the exterior power of the vertical bundle $\Psi: \Nt \rightarrow \wedge^2
T(\Nt/M)$.
Since we are interested in a local 
description, we shall avoid the development of  the usual infinite-dimensional formalism via the
introduction of a Poisson bracket as a local bilinear operator $\Psi(\Svec(x)) \delta(x - y)$ on the loop algebra of
state variations (see \cite{faddeev+takhtajan}), and instead immediately write the equations of motion in the local form for any point $x
\in M(t)$:
\begin{equation}
  \label{hml.eqmot}
   \Svec_t(x) =  \{H,\,\Svec\}(x) \equiv \sum_{a,b} \Psi_{a,b}(S(x))\vder{\Svec}{S_a}(x) \vder{H}{S_b}(x) = \Svec(x) \times \Svec_{xx}(x),
\end{equation}
where as  functions in $A_0(t)$ the coordinate functions $\Svec$ have the natural variational derivatives as defined in 
Section~\ref{Section.Setting}: $ \vder{S_a}{\Svec} = \sum_b \delta_{a,b} \delta S_b$.

To restore independence of the fiber coordinates in (\ref{hml.eqmot}) we can state that the 
vector product in (\ref{hml.eqmot}) is defined with respect to the standard metric on $\R^3$. 
There is a more natural way, however: the bracket (\ref{hml.bracket}) is the Lie-Poisson bracket on the space $\sothree*
\cong \R^3$ dual to the Lie algebra $\sothree$ 
\cite{karasev+maslov}, and therefore variational derivatives of functions in $\Ainf(t) = \Omega^{0,0} \Ninf(t)$ lie in
$\Omega_0^{1,0} \Ninf(t) \cong \Ainf(t) \otimes T^*\sothree^* \cong \Ainf(t) \otimes \sothree^*$, on which the bivector field
$\Psi \in \Ainf(t) \otimes \wedge^2 T\sothree^*$ acts naturally and independently of any
coordinate system.

The bracket (\ref{hml.bracket}) 
is nondegenerate on the leaves of a $2-$dimensional foliation of $\R^3$ consisting of 
spheres about the origin, orbits of the coadjoint action of $\text{SO}(3)$ \cite{karasev+maslov}.  
On each such leaf a symplectic form dual to the bracket, the {\em Kirillov form}, is defined.
Since, as is easily verified, the dynamics of
(\ref{hml.eqmot}) preserves the leaves \mbox{($|\Svec|^2_t = 0$)}, the system can be considered 
on the bundle $N(t) = M^{n-1} \times S^2$ with the unit sphere as the fiber, provided the initial conditions are chosen
as a section of this new bundle.  It is well-known that the symplectic form dual to $\Psi$ is the area form on the
sphere, which can be written down as soon as local coordinates have been chosen on $S^2$.  For instance,
the unit sphere with a deleted north pole can be parametrized by 
the inverse stereographic projection from the pole $(0,0,1)$.
Following \cite{faddeev+takhtajan} we introduce the following transformation to complex coordinates:
\begin{gather*}
  w = w_1 + i w_2,\ S = \Sone + i \Stwo,\ \ Q = \Sthree, \ R = \frac{1 + \wsq}{2} > 0,\quad
  \bS = \left(S,\,\Sb,\, Q\right)^T,\quad
  \bW = \left(w,\,\wb\right)^T,
\end{gather*}
the correspondence is then established as follows:
\begin{equation}
        \label{hml.StoW}
        \begin{gathered}
                S = \frac{2 w}{1 + \wsq} = \frac{w}{R},\quad 
                \Sb = \frac{2 \wb}{1 + \wsq} = \frac{\wb}{R},\quad 
                Q = \frac{\wsq - 1}{\wsq + 1} = \frac{R-1}{R},\\
                \wsq = \frac{1 + Q}{1 - Q},\ R = \frac{1}{1 - Q},\ 
                w = \frac{S}{1 - Q} = S R,\ \wb = \Sb R.
        \end{gathered}
\end{equation}.

The Poisson brackets of the modified coordinate functions $\bS = \left(S,\,\Sb,\,Q\right)$ and the Hamiltonian density 
\begin{equation}
  \label{SQbracketHam}
  \{S,\,\Sb\} = = 2i Q,\quad
   \{S,\,Q\} =  -i S,\quad
   \{\Sb,\,Q\} = i \Sb,\quad
  H = \half \left(|S_x|^2 + Q_x^2 \right),
\end{equation}
generate the equations of motion equivalent to (\ref{hml.eqmot}).
The coordinate transformations (\ref{hml.StoW})
generate the bracket of the $\bW$-coordinates:
\begin{gather*}
  \label{Wbracket}
  \{w,\,\wb\} = R^2 \{S,\,\Sb\} + \Sb R^3 \{S,\,Q\} + S R^3 \{Q,\,\Sb\} = 
   -2i R^2.
\end{gather*}
This  bracket is clearly non-degenerate and defines a symplectic (Kirillov) form 
\begin{equation}
  \label{omega}
  \omega = \frac{\dw \W \dwb}{2i R^2} = \frac{2 \dw  \W \dwb}{i (1 + \wsq)^2},
\end{equation}
which after scaling $\displaystyle{ z = \frac{w}{\sqrt{R}}}$ 
\cite{faddeev+takhtajan} reduces to the canonical form 
$\dsp{\frac{\dz \W \dzb}{i}}$.
Since exterior differentiation commutes with pullback, we obtain a primitive for $\omega$
by pulling back a primitive $\theta$ of the canonical form:
\begin{gather*}
    \label{theta}
    \theta  = \frac{z \dzb - \zb \dz}{2i} =  
    \frac{1}{2i} \left[
           \left( \frac{w}{R} - \frac{w \wsq}{4 R^2}\right) \dwb 
           \ -\ \frac{\wb \wsq}{4 {R^2}} \dw
           \ -\ \left( \frac{\wb}{R} - \frac{\wb \wsq}{4 {R^2}}\right) \dw  
           + \frac{w \wsq}{4 {R^2}} \dwb
           \right] = 
    \frac{w \dwb - \wb \dw}{2i R}.
\end{gather*}
The density in (\ref{SQbracketHam}) transforms to 
\begin{equation}
  \label{WHam}
  H = \frac{\wx \wbx}{2 R^2},
\end{equation}
which generates equations of motion in the new bracket
\begin{equation}
        \wt = i\frac{-R \wxx + \wb \wx^2}{R} = i\frac{-\wxx - \wsq \wxx + 2 \wb \wx^2}{2R} = -i \wxx + i \frac{\wb \wx^2}{R}.
  \label{hm.eqmot_W}
\end{equation}

\subsection{Lagrangian and Multisymplectic formulation of HM}
We now have all the  elements necessary to reconstruct the Lagrangian and multisymplectic
form of the system in the complex 
plane.  With respect to the symplectic structure (\ref{omega}) the equations of motion (\ref{hm.eqmot_W}) are
$$
  \inner{\bW_t}{\omega} = \vder{H}{\bW}.
$$
From (\ref{setting.lagr}), letting $\theta$ be defined as in (\ref{theta}), 
$X_t = \tilde \partial_t = \wt \pder{}{w} + \wbt
\pder{}{\wb}$, and $\bH = H dx$ using (\ref{WHam}), one obtains 
\begin{equation}
  \label{LagrHM}
  \bLam = 
      -\left(H + \inner{X_t}{\theta}\right) dx\wedge dt = 
      \underbrace{-\left(\frac{\wx \wbx}{2 R^2} + \frac{w \wbt - \wb \wt}{2i R} \right) dx \wedge dt}_{\Lambda}.
\end{equation}
The multisymplectic form now follows from (\ref{setting.cov}).
Computing the variation of the Lagrangian one obtains
\begin{multline}
        \label{HMdLam}
        \delta \bLam = 
                       \left\{\frac{\wb \wx \wbx}{2 R^3} + \partial_x \left( \frac{\wbx}{2 R^2}\right)
                              - \frac{\wbt}{2i R} + \frac{w \wb \wbt - \wb^2 \wt}{4iR^2}
                              - \partial_t \left( \frac{\wb}{2i R} \right) 
                       \right\} \dw \W dx \W dt \\
        \ + \ \\                     
                       \left\{\frac{w \wx \wbx}{2 R^3} + \partial_x \left( \frac{\wx}{2 R^2} \right)
                              + \frac{\wt}{2i R} + \frac{w^2 \wbt - w \wb \wt}{4iR^2}
                              + \partial_t \left( \frac{w}{2i R} \right) 
                       \right\} \dwb \W dx \W dt \\        
        \ + \ 
        d \left\{  \frac{\wbx}{2 R^2} \dw \W dt + \frac{\wx}{2R^2} \dwb \W dt + 
                   \frac{\wb}{2iR} \dw \W dx  - \frac{w}{2iR} \dwb \W dx \right\},
\end{multline}
yielding the fundamental differential forms
\begin{equation}
        \label{HMOm1}
        \Theta = -\frac{\wbx}{2 R^2} \dw \W dt - \frac{\wx}{2 R^2} \dwb \W dt - 
                  \frac{\wb}{2i R} \dw \W dx + \frac{w}{2iR} \dwb \W dx,
\end{equation}
\begin{multline}
        \label{HMOm}
        \Om = \delta \Theta \frac{\wb \wx - w \wbx}{2 R^3} \dw \W \dwb \W dt + 
              \frac{1}{2R^2} \dw \W \dwbx \W dt + \frac{1}{2R^2} \dwb \W \dwx \W dt
        \ +\ \frac{1}{2iR^2} \dw \W \dwb \W dx,
\end{multline}
and, by using (\ref{setting.cov}), the covariant Hamiltonian
\begin{equation}
        \bHam = 
               \left(\frac{\wx \wbx}{2 R^2} + \frac{w \wbt - \wb \wt}{2i R} \right)dx \W dt + 
               \left( \frac{\wb \wt - w \wbt}{2i R} -\frac{\wx \wbx}{R^2} \right) dx \W dt = 
                -\frac{\wx \wbx}{2R^2} dx \W dt = -\bH.
\end{equation}
Comparing the coefficients of $\delta w \W dx \W dt$ and $\delta \overline w \W dx \W dt$ in
\begin{equation*}
        \delta \bHam = 
        \frac{\wb \wx \wbx}{2 R^3} \dw \W dx \W dt\ +\ \frac{\w \wx \wbx}{2R^3} \dwb \W dx \W dt
        \  -\ \frac{\wbx}{2R^2} \dwx \W dx \W dt\ -\ \frac{\wx}{2R^2} \dwbx \W dx \W dt,
\end{equation*}
to those in
\begin{multline*}
        dx \W \inner{\tilde \partial_x} \Om = 
               \frac{R \wxx + \w \wx \wbx - \wb \wx^2}{2 R^3} \dwb \W dx \W dt\ +\  
               \frac{R \wbxx + \wb \wx \wbx - \w \wbx^2}{2 R^3} \dw \W dx \W dt  \\
            \  -\ \frac{\wx}{2R^2} \dwbx \W dx \W dt - \frac{\wbx}{2R^2} \dwx \W dx \W dt,\quad
        dt \W \inner{\tilde \partial_t} \Om = 
               \frac{\wt}{2iR^2} \dwb \W dx \W dt - \frac{\wbt}{2iR^2} \dw \W dx \W dt,
\end{multline*}
we obtain the equations of motion (\ref{hm.eqmot_W}), while the coefficients 
of $\dwx \W dx \W dt$ and $\dwbx \W dx \W dt$ yield the tautological identities
$$
        -\frac{\wbx}{2 R^2} = -\frac{\wbx}{2 R^2},\quad
        -\frac{\wx}{2 R^2} = -\frac{\wx}{2 R^2}.
$$
These can be used to define momenta as in (\ref{setting.momenta}):
\begin{gather}
  \p \equiv P_{1,(1,0)} = \vder{\Lam}{\wx} = \pder{\Lam}{\wx} = -\pder{H}{\wx} = -\frac{\wbx}{2R^2}, \quad
  \pb \equiv P_{2,(1,0)} = \vder{\Lam}{\wbx} = \pder{\Lam}{\wbx} = -\pder{H}{\wbx} = -\frac{\wx}{2R^2},
\end{gather}
so that in new fiber coordinates $(w,\,\wb,\,p,\,\pb,\,\wt,\,\wbt)$ we have
\begin{gather}
  \bHam = - 2 R^2 p \pb dx \W dt = -(1 + \wsq) \psq dx \W dt, \label{HMHamwp}\quad
  \Om = - \dw \W \delp \W dt - \dwb \W \delpb \W dt + \frac{1}{2iR^2} \dw \W \dwb \W dx,
\end{gather}
and the components of (\ref{setting.cov}) take the form
$$
   dx \W \inner{\tilde \partial_x}{\Om} = \wx \delp \W dx \W dt - \px \dw \W dx \W dt 
      \ +\ \wbx \delpb \W dx \W dt - \pbx \dwb \W dx \W dt,
$$
$$
   dt \W \inner{\tilde \partial_t}{\Om} = \frac{\wt}{2i R^2} \dwb \W dx \W dt - 
   \frac{\wbt}{2i R^2}\dw \W dx \W dt,
$$
$$
   \delta \bHam = - 2 R \wb \psq \dw \W dx \W dt - 2 R \w \psq \dwb \W dx \W dt 
   - 2 R^2 \pb \delp \W dx \W dt - 2 R^2 p \delpb \W dx \W dt,
$$
yielding the equations
\begin{gather}
  - 2 R^2 \pb = \wx,\quad
  - 2 R^2 p = \wbx,\quad
  - 2 R \wb \psq = -\frac{\wbt}{2iR^2} - \px,\quad
   - 2 R \w \psq = \frac{\wt}{2i R^2} - \pbx,
\end{gather}
which are equivalent to (\ref{hm.eqmot_W}).  This is the local multisymplectic formulation of HM.

To relate our formalism to \cite{bridges+reich.ms} we introduce two matrices $\bK$ and $\bM$ and a coordinate vector 
$\bz = (w,\,\wb,\,p,\,\pb)$ (since $\wt,\,\wbt$ never show up explicitly, we can get away with fewer fiber coordinates):
\begin{equation}
  \bK = \begin{pmatrix}
           0 & 0 & -1 &  0 \\
           0 & 0 & 0  & -1 \\
           1 & 0 & 0  &  0 \\
           0 & 1 & 0  &  0           
        \end{pmatrix}, \quad
  \bM = \frac{1}{2iR^2}
        \begin{pmatrix}
           0 & -1 & 0  &  0 \\
           1 & 0  & 0  &  0 \\
           0 & 0  & 0  &  0 \\
           0 & 0  & 0  &  0           
        \end{pmatrix}.
\end{equation}
If $\Om = \Om_t dt + \Om_x dx$, where $\Om_t,\, \Om_x \in \Om^{(2,0)}$ are fiber $2$-forms,
then $\bK$ and $\bM$ are simply matrices for $\Om_t$ and $-\Om_x$ in $\bz$-coordinates.
Indeed, 
\begin{gather*}
  dx \W \inner{\tilde \partial_x}{\Om} = -\inner{\tilde \partial_x}{\Om_t} dx \W dt = 
  -\bz_x^T \bK \cdot \delta \bz \W dx \W dt,\quad
  dt \W \inner{\tilde \partial_t}{\Om} = \inner{\tilde \partial_t}{\Om_x} dx \W dt = 
  -\bz_t^T \bM \cdot \delta \bz \W dx \W dt,
\end{gather*}
and the canonical equations (\ref{setting.cov}) reduce to 
\begin{equation}
   \bK \cdot \bz_x + \bM \cdot \bz_t = \nabla_{\bz} \Ham,
\end{equation}
where $\nabla_{\bz} \Ham$ is a column vector of partial derivatives.

Finally, the conservation laws (\ref{setting.conserv}) yield the usual energy ($H$) and momentum ($P$) conservation laws:
\begin{gather}
        T_{tt} = \frac{w_x \wb_x}{2R^2} = -H,\ T_{tx} = -\frac{w_t \wb_x + \wb_t w_x}{2 R^2},\\
        T_{xt} = \frac{\wb w_x - w \wb_x}{2i R} = P,\ T_{xx} = -\frac{w_x \wb_x}{2 R^2} + \frac{w \wb_t - \wb w_t}{2i R}.
\end{gather}
It is important that the definition of these quantities ultimately relies on the choice of the primitive form $\theta$,
which is defined only locally.  In general, we have no global definition for the conserved densities $T_{jk}$ nor for
the Lagrangian density itself.  The functionals defined by these densities are known as {\em multivalued functionals}
since changes in local choices of $\theta$ result in differences by a multiple of a topological term, much like the multivalued function $\ln z$.  
This phenomenon was investigated, among others, in \cite{novikov, witten}, where \cite{novikov} sets a general task of
relating the properties of critical points of such functionals to
the topology of the underlying infinite-dimensional manifold of sections, 
extending the classical Morse theory, and,
similarly, in the finite-dimensional case $F$, replacing functions of the Morse theory by nontrivial elements $H^1(F, \R)$,
i.e. close but non-exact forms.  In our situation the finite-dimensional case ($\theta$) fully determines the
infinite-dimensional counterpart.  The essential idea of \cite{novikov}
is to find a covering manifold on which the $1$-form becomes exact, defining a global function, to which the classical 
Morse theory applies.  At the moment it is not clear to us what covering of the sphere $S^2$ will trivialize $\theta$.
Such a covering is unlikely to be obtained by reduction of the 
symmetry groups $SO(3)$ and $SU(2)$ since, being
$3$-dimensional they do not carry a natural symplectic structure themselves, and being compact they do not have
factor-spaces $\hat F$ with a nontrivial $H^2(\hat F, \R)$.  The necessary covering must be similar to an
infinitely-sheeted surface of $\ln z$, except covering the sphere completely (no branch points excluded).
Further, the classical theory due to Lyusternic-Shnirelmann extends the Morse theory in the infinite-dimensional case to
allow degenerate functionals, and relates
 the numbers of its critical points of different index to a topological
invariant called the {\em Lyusternik-Shnirelmann category} (see 
the recent work by Walter Craig on water waves).  This
theory does require a globally-defined Lagrangian function, which the Novikov-Morse theory promises to provide.  Given
this, we should be able to estimate the number of solutions of a given kind (say, periodic ones) of HM on a given domain
(e.g., with periodic boundary conditions) including the higher-dimensional spatial cases.  Similar considerations apply
to linearizations of HM near fixed points or periodic solutions, promising 
to be a tool of stability analysis.
These investigations will be continued elsewhere.

\section{HM discretizations}
\label{Section.HM-Discr}

Given a PDE in variational form there are approaches to its discretizations that more or less automatically produce
discrete analogs of the preservation of the multisymplectic form (\ref{setting.ms}), and frequently result in superior
discrete conservation laws (Cf. (\ref{setting.conserv}) which suggests a discrete analog of N\"other's theorem).
While \cite{marsden+patrick+shkoller} developed a finite-difference approach to variational PDEs,
we believe a more
natural approach is to rely on the finite element method (FEM) involving space-time meshes.  
One of the reasons is that the
equation already possesses a natural variational form,  unlike many equations treated by a version of Galerkin's method.
Here we outline our FEM-based approach to discretization of the HM model.

\subsection{Finite element method for HM}
Using (\ref{LagrHM}) the continuous action functional is (with ${\mathcal L} \equiv \tbLam$ as defined in (\ref{caf}))
$$
  {\mathcal L} = \int \bLam = \int \left(i \frac{w \wbt - \wb \wt}{2 R} - \frac{\wx \wbx}{2 R^2}\right) dx \W dt.
$$
Introducing a rectangular space-time mesh $\hat M$ on $M$ with elements $\{e\}$,
we take a basis of piecewise linear on $e$
functions $P^1$ (more generally they  can be piecewise polynomial) \cite{strang+fix}  which, on the canonical square
element with coordinates $(xi, \eta) \in [0,1] \times [0,1]$,  have the form
\begin{gather*}
     \phi_1 = \fourth (1 - \xi)(1-\eta), \quad \phi_2 = \fourth (1 + \xi)(1-\eta),\quad
     \phi_3 = \fourth (1 + \xi)(1+\eta), \quad \phi_4 = \fourth (1 - \xi)(1+\eta).
\end{gather*}
Then the approximations $w \doteq \sum_v w^v \phi_v$ are defined by collocation at element nodes (1 per element vertex, 4
per element), and for polynomial terms of $\mathcal L$ they reduce to polynomials in the nodal values $w^v$, while
the rational terms have no canonical representation.  First, we consider the {\em product approximation} 
\cite{christie+.prod_app}
$\frac{1}{R} \doteq r^v \phi_v = \frac{1}{R^v} \phi_v$,
and obtain the following
{\em discrete action functional}:
\begin{multline*}
  L = \frac{1}{2} \int\biggl[
    \sum_{u,v,\vb}    i r^u \phi_u w^v \wb^\vb \left(\phi_v\, \partial_t\phi_\vb - \partial_t\phi_v\, \phi_\vb\right)
    \ -\ \sum_{u,\ub,v,\vb}\left(r^u r^\ub w^v \wb^\vb \phi_u \phi_\ub \partial_x \phi_v\,\partial_x \phi_\vb \right)
                       \biggr]\ dx \W dt  = \\
 \sum_e \biggl[
 \sum_{a,\ab,b=1}^4 w_e^a \wb_e^\ab r_e^b 
 i \underbrace{\half\int_e \left(\phi_{e,a}\, \partial_t \phi_{e,\ab} - 
             \partial_t \phi_{e,a}\,\phi_{e,\ab}\right) \phi_{e,b}\ dx \W dt }_{A^e_{a,\ab,b}}
- \sum_{a,\ab,b,\bb=1}^4 w_e^a \wb_e^\ab r_e^b r_e^\bb \\
 \underbrace{\frac{1}{2}\int_e \partial_x \phi_{e,a}\, \partial_x 
              \phi_{e,\ab} \phi_{e,b} \phi_{e,\bb}\ dx \W dt}_{H^e_{a,\ab,b,\bb}} 
                         \biggr],
\end{multline*}
composed of {\em element functionals} $L_e$:
\begin{equation*}
  \label{DLagrHM-P}
  L = \sum_e L_e(w_e,\wb_e),\quad
  L_e(w_e,\wb_e) =  i\underbrace{\sum_{a,\ab,b=1}^4 w_e^a \wb_e^\ab r_e^b A^e_{a,\ab,b}}_{A_e(w_e,\wb_e)} 
               - \underbrace{\sum_{a,\ab,b,\bb=1}^4 w_e^a w_e^\ab r_e^b r_e^\bb H^e_{a,\ab,b,\bb}}_{H_e(w_e,\wb_e)},
\end{equation*}
called {\em element Lagrangians}.
The coefficients $A^e_{a,\ab,b}$ and $H^e_{a,\ab,b,\bb}$ are independent of the element $e$ taken to be a rectangle with
sides $h_x,\,h_t$, and which can be computed 
by reduction to the canonical integrals:
\begin{gather*}
 A^e_{a,\ab,b} = \hat A_{a,\ab,b} = \half \frac{h_x}{2} \int_{\hat e} 
   \left(\hat \phi_a\, \partial_\eta \hat \phi_\ab - \partial_\eta \hat \phi_a\, \hat \phi_\ab\right) 
     \hat \phi_b\ d\xi \W d\eta, \quad
 H^e_{a,\ab,b,\bb} = \hat H_{a,\ab,b,\bb} = \half \frac{h_t}{h_x} \int_{\hat e} 
   \partial_\xi \hat \phi_a\, \partial_\xi \hat \phi_\ab\, \hat \phi_b\, \hat \phi_\bb\ d\xi \W d\eta,
\end{gather*}
employing  the obvious symmetries:
$$
   \hat A_{a,\ab,b} = - \hat A_{\ab,a,b,\bb} \in \R,\quad 
   \hat H_{a,\ab,b,\bb} = \hat H_{\ab,a,b,\bb} = \hat H_{a,\ab,\bb,b} \in \R.
$$
Being polynomials of degree $\leq 8$ in ${\xi, \eta}$ these can be computed exactly (to round-off) 
using $5$-point Gaussian quadrature, and as in any FEM code are never derived explicitely but assembled at run-time.
While $\hat H$ is a full $4$-th order tensor, the $3$-rd order tensor $\hat A$ has the following sparsity structure:
$$
   \hat A_{\cdot,\cdot,b} \rightarrow 
     \begin{pmatrix} 
         0 & 0 & \times & \times\\
         0 & 0 & \times & \times\\
         \times & \times & 0 & 0\\
         \times & \times & 0 & 0\\ 
     \end{pmatrix}.
$$

As an alternative to the product approximation, we can discretize the rational terms by using their element averages.
The element average of a function is defined to be the average of its approximation 
in the space spanned by $\{\phi_v\}$:
$$
  \tilde f_e = \frac{1}{|e|} \int_e \sum_v f^v \phi_{v} = \frac{1}{|e|} \int_e \sum_{a=1}^4 f^a_e \phi_{e,a} =  
  \sum_{a=1}^4 f^a_e \frac{1}{|e|}\int_{\hat e} \hat \phi_a = \sum_a f^a_e \tilde {\hat \phi_a},\quad 
  |e| = \int_e 1 = h_x h_t,
$$
$f_e^a$ denotes the nodal value of $f$ at the $a$-th vertex of element $e$.
Computing the element averages easily yields: $\tilde{\hat \phi}_a = \frac{1}{4}, a = 1,\dots, 4$;
then
$$
  \tilde f_{e} = \frac{1}{4} (f_e^1 + f_e^2 + f_e^3 + f_e^4).
$$
Defining element moment matrices $\hat K^{(0)},\, \hat K^{(1)}, \hat K^{(2)}$:
\begin{gather}
  \hat K^{(0)}_{a,\ab} = \int \hat \phi_{a} \hat \phi_{\ab},\quad
  \hat K^{(1)}_{a,\ab} = \int \phi_a\, \partial_t \phi_\ab, \quad 
  \hat K^{(2)}_{a,\ab} = \int \partial_x \hat \phi_a\, \partial_x \hat \phi_{\ab},
\end{gather}
we have
\begin{gather}
  \hat K^{(0)} = \frac{h_x h_t}{36} \begin{pmatrix}
            4          & 2           & 2           & 1          \\
            2          & 4           & 1           & 2          \\
            2          & 1           & 4           & 2          \\
            1          & 2           & 2           & 4 
      \end{pmatrix},\quad
  \hat K^{(1)} = \frac{h_x}{12} \begin{pmatrix}
                -   2           &   -1           &\ \ 2           &\ \ 1          \\
                -   1           &   -2           &\ \ 1           &\ \ 2          \\
                -   1           &   -2           &\ \ 1           &\ \ 2          \\
                -   2           &   -1           &\ \ 2           &\ \ 1             
      \end{pmatrix},\quad
  \hat K^{(2)} = \frac{h_t}{6 h_x} \begin{pmatrix}
                \ \ 2          &   -2          &\ \ 1          &   -1          \\
                -   2          &\ \ 2          &   -1          &\ \ 1          \\
                \ \ 1          &   -1          &\ \ 2          &   -2          \\
                -   1          &\ \ 1          &   -2          &\ \ 2          
      \end{pmatrix}.
\end{gather}
Now substituting approximations $w \doteq \sum_v w^v \phi_v$ and $\Rt_e \doteq R(\tilde w_e)$ into $\mathcal L$
($\Rt$ is defined in the element interior only, which is enough for the integral to make sense),  
we obtain another
discrete action functional:
\begin{multline*}
  L = -\int \sum_{v,\vb}\left( \frac{w^v \wb^\vb \phi_v\, \phi_\vb}{2 \Rt^2} + 
                              \frac{w^v \wb^\vb (\phi_v \phi_{\vb,t} - \phi_{v,t} \phi_\vb)}{2 i \Rt}
                       \right)dx \W dt
   = \sum_e \underbrace{\sum_{a,\ab =1}^4 \left( -\frac{w_e^a \wb_e^\ab}{2 \Rt_e^2} \hat K^{(2)}_{a,\ab} -
                              \frac{w_e^a \wb_e^\ab}{2 i \Rt_e} 
                              (\hat K^{(1)}_{a,\ab} - \hat K^{(1)}_{\ab,a}) \right)}_{{L_e}(w,\wb)}.
\end{multline*}
Noting that
$$
  \hat K^{(1)} - ({\hat K^{(1)}})^T = \frac{h_x}{4}
                         \begin{pmatrix} 0 & I\\ -I & 0 \end{pmatrix} = \hat J,
$$
and letting $\hat K = \hat K^{(2)}$ we can write
\begin{equation}
  \label{DLagrHM-A}
  L = \sum_e L_e(w,\wb),\quad
  L_e(w,\wb) =  \sum_{a,\ab=1}^4 \frac{w_e^a \wb_e^\ab}{2 \Rt_e} 
                                     \left( i \hat J_{a,\ab} - \frac{\hat K_{a,\ab}}{\Rt_e}\right).
\end{equation}

The {\em element Lagrangian} $\hat L_e$ corresponds to $L_{\Delta}$ in \cite{marsden+patrick+shkoller}
and represents one element's contribution to the action.

For a regular mesh the discrete Lagrangian is invariant under a mesh shift and as can be easily seen from (\ref{DLagrHM-A}), 
then the element Lagrangian is defined in terms of the {\em canonical} element Lagrangian $\hat L$:
\begin{gather} 
   L_e(w,\wb) = \hat L(w_e^1,\, w_e^2,\, w_e^4,\, w_e^4),\quad
   \hat L(w^1,\, w^2,\, w^3,\, w^4) =  \sum_{a,\ab=1}^4 \frac{w^a \wb^\ab}{2 \Rt} 
                                     \left( i \hat J_{a,\ab} - \frac{\hat K_{a,\ab}}{\Rt}\right),
\end{gather} 
where $(w_e^1,\, w_e^2, w_e^3, w_e^4)$ denote the four nodal points at a particular element,
and $\Rt$ is defined using the average of the four values $w^1, w^2, w^3, w^4$.
Defining $\hat J(w,\,\wb)$ and $\hat K(w,\,\wb)$ by their matrices
\begin{gather*}
    \Jh(w,\,\wb) = w \cdot \Jh \cdot \wb = \sum_{a,\ab} w_a \Jh_{a,\ab} \wb_\ab,\quad
    \Kh(w,\,\wb) = w \cdot \Kh \cdot \wb = \sum_{a,\ab} w_a \Kh_{a,\ab} \wb_\ab,
\end{gather*}
we obtain
\begin{multline}
    \Lh(w,\,\wb) =  \frac{i}{2 \Rt} \Jh(w,\wb)  - \frac{1}{2\Rt^2} \Kh(w,\,\wb),
    \frac{i h_x}{2\Rt} \left[ \frac{(\wone \wthreeb - \woneb \wthree)}{4} + \frac{(\wtwo \wfourb - \wtwob \wfour)}{4}
                       \right] - \\
    \frac{1}{2\Rt^2} \frac{h_t}{6 h_x}
    \left[ 2|\wone|^2 + 2|\wtwo|^2 + 2|\wthree|^2 + 2|\wfour|^2  \right.\\
         - 2(\wone\wtwob+\woneb\wtwob) + (\wone \wthreeb + \woneb \wthree)
          - (\wone \wfourb + \woneb \wfour) - (\wtwo \wthreeb + \wtwob \wthree)\\
    \left.      + (\wtwo \wfourb + \wtwob \wfour) - 2(\wthree \wfourb + \wthreeb \wfour) \right].
\end{multline}

The discrete Euler-Lagrange field equations (DELF, \cite{marsden+patrick+shkoller}) are obtained 
by differentiating the discrete action, which
depends on a particular nodal value $w^v$ only through 
the element Lagrangians corresponding to four elements $e_v^a,\ a = 1, \dots, 4$ containing $v$ as one of their
vertices: $L_{e_v^1},\,L_{e_v^2},\,L_{e_v^3},\,L_{e_v^4}$.
DELF equations are written simply as
$$
   \pder{L}{w^v} = 0,\quad  \pder{L}{\wb^v} = 0,\quad
   \pder{\Rt}{w^a} = \frac{1}{8} \wbtil,
$$
determined using the canonical element derivative
$$
   \pder{\hat L}{w^a} = \frac{i}{2\Rt} \Jh_a \cdot \wb
                      -  \frac{i\wbtil}{16 \Rt^2} \Jh(w,\,\wb)
                      - \frac{1}{2 \Rt^2} \Kh_a \cdot \wb
                      +  \frac{\wbtil}{8\Rt^3} \Kh(w,\,\wb).
$$

\subsection{Alternative approaches}
\label{SubSection.HM-Disc-alt}
Preliminary numerical experiments indicate that the FEM-based methods developed above can be unstable, apparently due to
a relatively poor approximation of the rational terms in the polynomial bases.  
Here we sketch alternative approaches that may
help avoid these difficulties.

First, consider a local quasi-linear form of the HM equations of motion.
For this construction we rewrite the Lagrangian for HM and the equations of motion in the real coordinates
$w = a + ib$:
\begin{multline*}
\bLam = \int \left(\frac{\wt \wb - \w \wbt}{2 i R} -  \frac{\wx \wbx}{2R^2}\right) dx \W dt = 
        \int \left( Im\left(\frac{\wb \wt}{R}\right) - Re \left(\frac{\wx \wbx}{2R^2}\right) \right) dx \W dt = \\
        \int \left( \underbrace{\frac{a b_t - a_t b}{R}}_{A} - \underbrace{\frac{a_x^2 + b_x^2}{2R^2}}_{H} \right) dx \W dt.
\end{multline*}
Applying the Euler-Lagrange operator explicitly we obtain
\begin{gather*}
  \vder{A}{a} = \frac{b_t}{R} + \partial_t\left(\frac{b}{R}\right) - a\, \frac{a\,b_t - a_t\,b}{R^2} = 
  \frac{2\,b_t}{R} - \frac{b}{R^2}\,(a\,a_t + b\,b_t) - \frac{a^2 b_t} + \frac{a_t\,a\,b}{R^2} = 
  \frac{2\,b_t}{R} - \frac{(a^2+b^2)b_t}{R^2} = \frac{b_t}{R^2},\\
  \vder{H}{a} = \partial_x \left(\frac{a_x}{R^2}\right) + a\,\frac{a_x^2 + b_x^2}{R^3} = 
  \frac{a_{xx}}{R^2} - 2\,\frac{a_x}{R^3}\,(a\,a_x + b\,b_x) + a\,\frac{a_x^2 + b_x^2}{R^3} = 
  \frac{a_{xx}}{R^2} + \frac{a\,(b_x^2 - a_x^2) - 2 a_x\,b\,b_x}{R^3}.
\end{gather*}
Since $A$ is anti-symmetric in $a$ and $b$ while $H$ is symmetric in the same variables, we easily obtain 
the corresponding expressions for $\dsp{\vder{A}{b}}$ and $\dsp{\vder{H}{b}}$:
\begin{gather*}
  \vder{A}{b} = \frac{b_t}{R^2},\quad
  \vder{H}{b} = \frac{b_{xx}}{R^2} + \frac{b\,(a_x^2 - b_x^2) - 2 a\,a_x\,b_x}{R^3}.
\end{gather*}
Thus, the Euler-Lagrange equations are
\begin{gather}
  \vder{\bLam}{a} = \frac{b_t}{R^2} - \frac{a_{xx}}{R^2} + \frac{a(a_x^2 - b_x^2) + 2 a_x\, b\, b_x}{R^3} = 0,\quad
  \vder{\bLam}{b} = -\frac{a_t}{R^2} - \frac{b_{xx}}{R^2} + \frac{b(b_x^2 - a_x^2) + 2 a\, a_x\, b_x}{R^3} = 0,
\end{gather}
which are equivalent to (\ref{hm.eqmot_W}) or
\begin{gather}
  \partial_t a + \partial_x b_x + \frac{b\,(b_x^2 - a_x^2) + 2\,a\,a_x\,b_x}{R} = 0,\quad
  \partial_t b - \partial_x a_x + \frac{a\,(a_x^2 - b_x^2) + 2\,a_x\,b\,b_x}{R} = 0.
\end{gather}
This equation has the quasi-linear and even the semi-linear form and is amenable to the method of characteristics.

Another approach, relying on the construction of a special class of section bases spaces relies on Lie-group
methods.  Limitations of space allow us only to sketch the approach.  The group $\SOthree$ acts transitively on $S^2$,
therefore on each element $e$ with canonical local coordinates $e$ have $\Svec(x,t) = g(x,t) \cdot \Svec_0$, where $g
\in \SOthree$ and the action should be coadjoint 
(unlike the standard linear representation, found in the numerical
literature), since it interacts naturally with the commutator.  The group near the identity must be parameterized by its
Lie algebra $\cong \R^3$, so that $g(x,t) = \widehat \exp(x A(x,t) + t B(x,t))$ where $A$ and $B$ are polynomial functions
of $x,t$ with values in $\sothree$ and $\widehat \exp$ is an approximation to the exponential map (e.g. the Caley
transform $(I + \half A)(I - \half A)^{-1}$).  Using a convenient choice
of $A$ and $B$ we should be able to reduce the spin length constraint (orbit throught $\Svec_0$ diffeomorphic to a
the coset space by a close subgroup of rotations around $\Svec_0$), but is linearization (ideal in $\sothree$).  The
spaces of sections (no longer linear) generated by $A$ and $B$ in a certain smoothness class replace the traditional FEM
bases. The details are the subject of a forthcoming publication.

In this paper we have established the local Lagrangian and multisymplectic structure of the Heisenberg magnet model, and
have shown how the powerful formalism of the variational calculus can naturally lead to the conservation properties and
suggest natural discretizations.  

\begin{thebibliography}{10}

\bibitem{bridges+reich.ms}
T.~Bridges and S.~Reich.
\newblock Multi-symplectic integrators; numerical schemes for hamiltonian pdes
  that conserve symplecticity.
\newblock {\em Phys. Lett. A}, 284:184--193.

\bibitem{christie+.prod_app}
I.~Christie, D.~Griffiths, A.~Mitchell, and J.~Sanz-Serna.
\newblock Product approximation for nonlinear problems in the finite element
  method.
\newblock {\em IMA J. Num. Anal.}, 1:253--266, 1981.

\bibitem{dikii}
L.~Dickey.
\newblock {\em Soliton Equations and Hamiltonian Systems}.
\newblock World Scientific, 1987.

\bibitem{deligne+.I}
P.~Deligne et~al.
\newblock {\em Quantum Fields and Strings: A Course for Mathematicians}.
\newblock American Mathematical Society, Institute for Advanced Study, 1999.

\bibitem{faddeev+takhtajan}
L.~D. Faddeev and L.~A. Takhtajan.
\newblock {\em Hamiltonian Methods in Soliton Theory}.
\newblock Springer-Verlag, 1987.

\bibitem{karasev+maslov}
M.~V. Karasev and V.~P. Maslov.
\newblock {\em Nonlinear Poisson brackets. Geometry and Quantization}.
\newblock American Mathematical Society, 1993.

\bibitem{manin}
Y.~Manin.
\newblock Algebraic aspects of nonlinear differential equations.
\newblock {\em Itogi nauki i tekhniki}, 11:5--152, 1978.

\bibitem{marsden+patrick+shkoller}
G.~Marsden, J.~Patrick and S.~Shkoller.
\newblock Multisymplectic geometry, variational integrators and nonlinear
  {PDEs}.
\newblock {\em Communications in Mathematical Physics}, 199:351--395, 1998.

\bibitem{novikov}
S.~P. Novikov.
\newblock Hamiltonian formalism and a multivalued analog of morse theory.
\newblock {\em Russian Mathatical Surveys}, 37(5):3--49, 1982.

\bibitem{palais}
R.~Palais.
\newblock {\em Foundations of Global Nonlinear Analysis}.
\newblock W. A. Benjamin, INC., 1968.

\bibitem{steenrod}
N.~Steenrod.
\newblock {\em The Topology of Fibre Bundles}.
\newblock Princeton University Press, 1951.

\bibitem{strang+fix}
G.~Strang and G.~Fix.
\newblock {\em An Analysis of the Finite Element Method}.
\newblock Series in Automatic Computation. Prentice-Hall, 1973.

\bibitem{witten}
E.~Witten.
\newblock Global aspects of current algebra.
\newblock {\em Nucl. Phys.}, B223(2):422--432, 1983.

\end{thebibliography}

\end{document}